# Conversion of any finite bandwidth optical field into a shape invariant beam

V. Arrizón, F. Soto-Eguibar, D. Sánchez-de-la-Llave, and H. M. Moya-Cessa

*Instituto Nacional de Astrofísica, Óptica y Electrónica, Apdo. Postal 51 y 216, Puebla PUE 72000, México*

**Abstract:** We discuss a method to transform any optical field, with finite frequency bandwidth, into a shape invariant beam with transverse scaling, dependent on the propagation distance. The method consists in modulating the field with a quadratic phase of appropriate curvature radius. As a particular application, we employ the method to extend the existence region of a finite non-diffracting field to an unbounded domain.

## 1. Introduction

Propagation of a coherent or partially coherent light beam is an important process in applied and theoretical optics. Researchers have devoted special attention to light beams that preserve their intensity profile during propagation. The so called non-diffracting (ND) beams, as Bessel [1-4], Mathieu [5], and Weber [6] beams, maintain their transverse intensity profiles without a scale change. Unfortunately, in the practical implementation of these beams, which include a finite envelope [4, 7-10], the invariant structure is only present in a finite region. Other optical fields, which we refer to as scaled propagation invariant (SPI) beams, preserve their shapes in an unbounded propagation range, at the expense of a re-scaling in their transverse extensions. The Hermite-Gaussian, Laguerre-Gaussian beams are examples of such optical fields [11].

Interesting methods, intended to extend the existence domain of Bessel beams, have been previously reported [12-15]. These beams are modified in such a way that they also present a re-scaling in their transverse widths.

Here we discuss a method to transform any optical field, with finite frequency bandwidth, into a SPI beam. The principle of the method is quite simple. It consists in modulating the field with a quadratic phase of appropriate curvature radius. This phase modulation can be obtained (for example) by passing the original beam through a divergent lens. As a particular result, we will show that the method allows the extension of the existence region of a ND beam to an unlimited propagation range.

It is interesting to note that the modification of a beam by a quadratic phase may occur (as a particular case) in the generalized Airy-Gauss beams, proposed in Ref. [9]. However, the possible extension of the existence domain of the beam, due to the presence of the quadratic phase, was not analyzed in this cited work.

The theoretical support of our proposal is discussed in section 2. The relevance and implications of the proposal is illustrated by numerical examples in section 3. Finally, conclusions and remarks are presented in section 4.

## 2. Theory

To start our discussion, let us consider an optical field $f(x,y)$, defined at the plane $z=0$, and its Fourier transform $F(u,v)$. The Fraunhofer propagation of the field $f(x,y)$, to a distance $z_0$, is given by $h(x,y)=E(x,y,z_0) F(x/\lambda z_0, y/\lambda z_0)$ [16], where $\lambda$ is the field wave-length and

$$E(x, y, z_0) = \exp\left[ ik(x^2 + y^2)/(2z_0) \right] \qquad (1)$$

is a quadratic phase of curvature radius $z_0$ and wave number $k=2\pi/\lambda$. An interesting fact, little considered in optics literature, is that $h(x,y)$ is a field whose transverse shape is invariant for propagation to any distance $z>z_0$, with a $z$-dependent scale change. A crucial fact that we have noticed is that this invariance is possible by the presence of the quadratic phase factor in the complex amplitude of $h(x,y)$. This fact motivated us to analyze the propagation of an arbitrary optical field when it is modulated by a quadratic phase factor.

Let us consider a monochromatic field of wavelength $\lambda$ whose complex amplitude, at the plane $z=0$, is $g_0(x,y)$. We will study the phase modulated version of this field, given by

$$g(x,y) = E(x,y,R)g_0(x,y). \tag{2}$$

For the sake of compactness, in the following mathematical analysis, we omit some constant complex factors. To compute the propagation of $g(x,y)$ we first obtain its Fourier transform. Using the convolution theorem in Eq. (2), and defining the vectors $V=(u,v)$ and $V_0=(u_0,v_0)$, the Fourier transform of $g(x,y)$ is given by the integral

$$G(u,v) = \iint_\Omega G_0(u_0,v_0)\exp(-i\pi\lambda R|V-V_0|^2)du_0 dv_0, \tag{3}$$

where $G_0(u,v)$ is the Fourier transform of $g_0(x,y)$. From now on we consider that the non-zero values of $G_0(u,v)$ appear in a finite circle of radius $\rho_C$, denoted as $\Omega$, which corresponds to the integration domain in Eq. (3). On the other hand, we assume the restriction

$$\lambda R \rho_C^2 = A << 1. \tag{4}$$

Under such restrictions, imposed to $G_0(u,v)$, the exponential $\exp[-i\pi\lambda R(u_0^2+v_0^2)]$ that appears by developing the argument of the integral in Eq. (3), is approximated to 1; and this integral is solved as

$$G(u,v) = \exp\left[-i\pi\lambda R(u^2+v^2)\right]g_0(\lambda Ru, \lambda Rv). \tag{5}$$

From Eqs. (2) and (5), it is noticed that $g(x,y)$ and its Fourier transform $G(u,v)$ have similar structures, except by the phase conjugation in Eq. (5). Thus, the field $g(x,y)$ can be called quasi-self-Fourier function [17-19].

In the Fresnel approximation, the propagation of the optical field $g(x,y)$, to the plane $z=z_0$, is given by

$$g_p(x,y) = \mathfrak{I}^{-1}\{G(u,v)\exp[-i\pi\lambda z_0(u^2+v^2)]\}, \tag{6}$$

where $\mathfrak{I}^{-1}$ is the inverse Fourier transform operator. Considering Eqs. (5) and (6), the propagated field $g_p(x,y)$ can be expressed by means of $E(x,y,R+z_0)\otimes G_0(-x/\lambda R,-y/\lambda R)$ where $\otimes$ is the convolution operator and $E$ is the quadratic phase defined in Eq. (1). By developing the explicit integral form of this convolution, applying again the restrictions imposed to the Fourier spectrum $G_0$, we obtain the propagated field

$$g_p(x,y) = E(x,y,MR)g_0\left(\frac{x}{M}, \frac{y}{M}\right), \tag{7}$$

with the scaling factor

$$M = 1 + z_0/R. \tag{8}$$

According to Eq. (7), the propagated field $g_p(x,y)$ is identical to the field $g(x,y)$, except by a re-scaling in the coordinates of the factor $g_0(x,y)$ and the curvature radius of the divergent phase. The field $g_p(x,y)$, in Eq. (7), provides a good approximation to the exact propagated field if the curvature radius $R$ fulfills the condition in Eq. (4). According to this relationship the approximation improves by decreasing $A$ (or $R$).

Since the expression for the field in Eq. (7) corresponds to the Fresnel propagation of the field $g(x,y)$, the distance $z_0$ should be (in principle) large enough to satisfy this propagation regime. However, we can prove that (under the adopted assumptions) Eq. (7) is a good approximation for any positive propagation distance $z_0$. First, we note that the field $g(x,y)$ at $z=0$ [Eq. (2)] represents the Fraunhofer propagation of the field $G_0(-x/\lambda R, -y/\lambda R)$, when it is located at the plane $z=-R$ [Fig. 1]. For the validity of this result it is only required to fulfill the band-width restriction imposed to the function $G_0$. But in this context, it can be proved that the far field propagated from the plane $z=-R$ to the plane $z=z_0$ (for any distance $z_0 \geq 0$) is also given by Eq. (7). Therefore, Eq. (7) provides a good approximation for the propagation of the field $g(x,y)$ to the plane $z=z_0$, regardless the distance $z_0$.

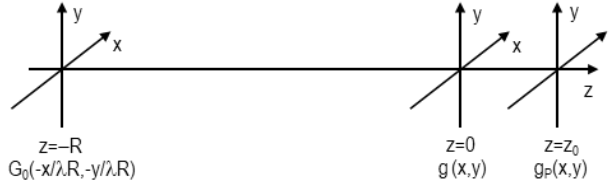

Fig. 1. Generation of the propagated fields in the far field domain of the field $G_0(-x/\lambda R,-y/\lambda R)$.

### 3. Numerical examples

To develop the first example we define the function

$$g_1(x) = \text{sinc}(x/p) \exp[-(x/w_0)^2], \tag{9}$$

where $w_0$ is the waist radius of the Gaussian factor, $p$ is the distance between adjacent roots of $g_1(x)$, and the 'sinc' function is defined as $\text{sinc}(\alpha)=(\pi\alpha)^{-1}\sin(\pi\alpha)$. Next, we define $g_0(x,y)= g_1(x)g_1(y)$, i. e.

$$g_0(x,y) = \text{sinc}\left(\frac{x}{p}\right)\text{sinc}\left(\frac{y}{p}\right)\exp\left(-\frac{x^2+y^2}{w_0^2}\right) \tag{10}$$

that we name as sinc-Gaussian (SG) field. The Fourier transform of $g_0(x,y)$ is $G_0(u,v)= G_1(u)G_1(v)$, where $G_1(u)=\text{rect}(pu)\otimes\exp(-\pi^2 w_0^2 u^2)$ and $\text{rect}(pu)$ is a rectangular pulse of width $1/p$ (in the $u$-domain). Now we assume the waist radius $w_0=6p$. In this case, the width of the Gaussian term, in the above convolution that defines $G_1(u)$, is smaller than the width of the rectangular pulse. Thus, the radius of the circular domain $\Omega$, which contains the non-zero values of $G_0(u,v)$ is approximately $\rho_C=(2)^{-1/2}p^{-1}$. In addition, assuming $A=1/20$ in Eq. (4), we obtain $R=p^2/(10\lambda)$. Thus, by employing Eq. (2), we obtain the complex amplitude of the phase modulated sinc-Gaussian (PMSG) field

$$g(x,y) = \exp[i 10\pi(x^2+y^2)/p^2]\, g_0(x,y). \tag{11}$$

The normalized transverse amplitude of the SG (or PMSG) field is displayed in Fig. 2 (a). In addition, the normalized transverse amplitudes of the Fourier spectra of the SG and PMSG fields are shown in Figs. 2 (b) and 2 (c). As expected, the Fourier spectrum of the SG field is a rectangular pulse (of width $1/p$), softened by the convolution with a narrow Gaussian function. On the other hand, the amplitude shape in Fig. 2 (c) confirms the striking result in Eq. (5), i. e. the Fourier spectrum $G(u,v)$ inherits the shape of $g_0(x,y)$.

According to Eq. (7), it is also expected that the propagated PMSG field inherits the transverse shape of the SG field, for any propagation distance $z_0>0$. To verify it we computed the propagated PMSG beam, using the exact propagation operator. The amplitude of this beam, for $z_0$ in the range $[0, 2R]$, is depicted in Fig. 3. In this result, the linear dependence of the transverse magnification, respect to $z_0$, is in agreement with Eq. (8). On the other hand, the field attenuation for $z_0>0$ is due to the normalization of the computed amplitudes, respect to the peak amplitude at $z_0=0$.

In the example under discussion, the curvature radius $R$ was computed with Eq. (4) using $A=1/20$, which is much smaller than 1. Under this condition, the propagated PMSG beam preserve the shape of the SG field $g_0(x,y)$, with good approximation. Even when we depicted the field amplitudes (instead of intensities), the errors of the field shapes in Figs. 2 (c) and 3, respect to the amplitude of the field $g_0(x,y)$, are not distinguished. Such errors are more significant when the restriction $A\ll 1$ [Eq. (4)] is transgressed. This situation is considered in the next example.

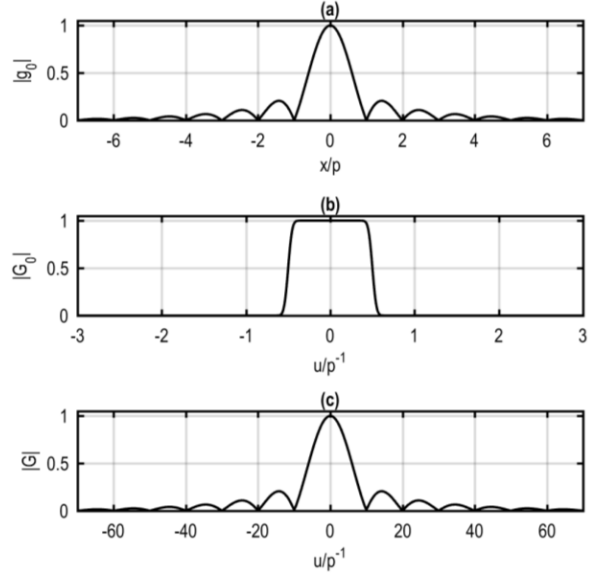

Fig. 2. (a) Transverse amplitudes of the SG field in Eq. (10) and (b) its Fourier spectrum. In (c) we show the transverse amplitude of the Fourier transform of the PMSG field defined in Eq. (11).

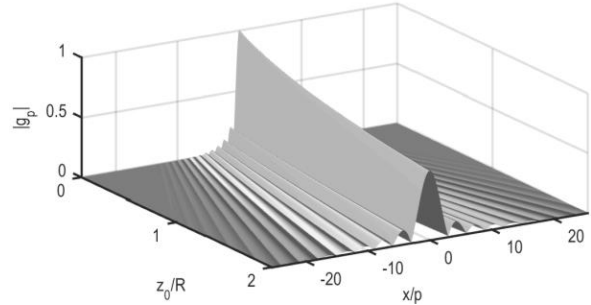

Fig. 3. Transverse amplitude of the propagated PMSG field, at the range $[0, 2R]$ for $z_0$.

As second case we consider the Bessel-Gaussian (BG) beam

$$g_0(x,y) = J_0(2\pi\rho_0 r)\exp(-r^2/w_0^2), \qquad (12)$$

where r is the radial coordinate, $\rho_0$ is the asymptotic radial frequency, and $w_0$ is the waist radius. By making the reasonable assumption $w_0 \gg \rho_0^{-1}$ (where $\rho_0^{-1}$ is the asymptotic beam period), we establish that the radius ($\rho_C$) of the circle that contains the non-zero values of the beam Fourier spectrum $G_0(u,v)$ is approximately $\rho_0$. Using this result in Eqs. (2) and (4), we obtain the phase modulated BG (PMBG) beam

$$g(x,y) = \exp(i\pi\rho_0^2 r^2/A)\, g_0(x,y). \qquad (13)$$

Let us consider specific BG and PMBG beams with the parameters $w_0=3\rho_0^{-1}$ and $A=1/4$. The amplitude of the BG (or PMBG) beam, at $z=0$, is shown in Fig. 4 (a); and the amplitudes of the normalized Fourier spectra $G_0(u,v)$ and $G(u,v)$ are depicted in Figs. 4 (b) and 4 (c). The transverse amplitude of $G_0(u,v)$ corresponds, as expected, to an annular field of radius $\rho_0$, whose transverse section is a Gaussian; and the plot in Fig. 4 (c) confirms that the Fourier spectrum $G(u,v)$ inherits the shape of $g_0(x,y)$.

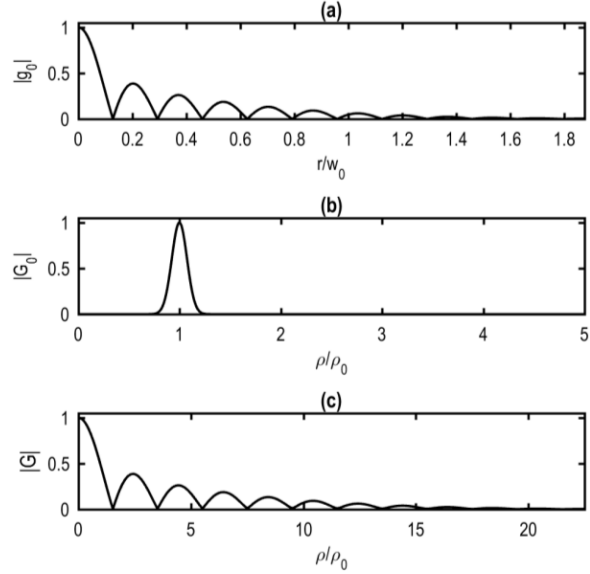

Fig. 4. (a) Transverse amplitudes of the BG field in Eq. (12) and (b) its Fourier spectrum. In (c) we show the transverse amplitude of the Fourier transform of the PMBG field $g(x,y)$ obtained with $A=1/4$.

Now we establish the longitudinal size of the existence zone of the BG beam, roughly given by $z_C=w_0/\lambda\rho_0$. This length is obtained geometrically, assuming that the beam radius is $w_0$. We computed the propagation of the BG and PMBG beams, for $z_0$ at the range $[0, 1.5z_C]$, using exact propagation. The amplitudes of the propagated fields $g_{0p}(x,y)$ and $g_p(x,y)$ in this range are depicted in Figs. 5 (a) and 5 (b). To establish the transverse scale in Fig. 5 we introduced the asymptotic BG beam period $p=\rho_0^{-1}$. The transverse amplitudes of the propagated BG beam $g_{0p}(x,y)$, at the distances $z_0=0$, $0.75\,z_C$ and $1.5\,z_C$, are displayed in Fig. 5 (c). In this figure, the transverse amplitude at the distance $z_0=1.5\,z_C$ (black line) tends to adopt the annular form of the BG beam far field. The transverse amplitudes of the propagated PMBG beam, at the distances $z_0=0$, $0.12\,z_C$, and $0.75\,z_C$, are displayed in Fig. 5 (d). The beams in Figs. 5 (b) and 5 (d), which results from the propagation of the PMBG beam, preserves the amplitude shape of this field. The expected transverse scaling is not visible in these figures because the transverse coordinate $r$ has been normalized respect to the length $M(z_0)p$, where $M(z_0)$ is the scale factor defined in Eq. (8). In Fig. 6 the transverse scaling of the propagated PMBG beam is made visible by the adopting the normalized transverse coordinate $r/p$ (independent of the propagation distance). In this case, the propagation distance $z_0$ is limited to the range $[0, z_C/4]$.

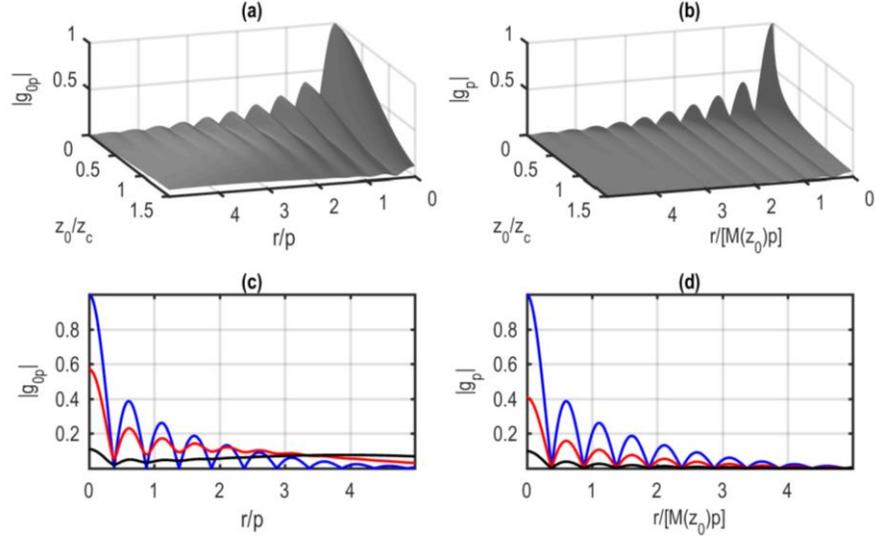

Fig. 5. Transverse amplitudes of the propagated (a) BG beam and (b) PMBG beam (for $A=1/4$) at the propagation range $[0,1.5z_C]$. The transverse amplitudes at three specific propagation distances for the BG beam and the PMBG beam are displayed in (c, d), respectively.

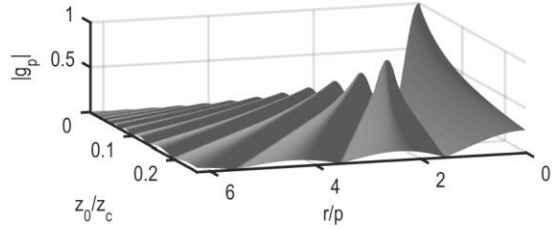

Fig. 6. Transverse amplitude of the propagated PMBG beam (for $A=1/4$) at the propagation range $[0, z_C/4]$. In this case the transverse coordinate normalization is the same for every $z_0$.

The arguments employed to obtain the PMBG beam in Eq. (13), can also be applied if the Bessel function $J_0(2\pi\rho_0 r)$ is replaced by any ND beam with Fourier spectrum $B(\phi)\,\delta(\rho-\rho_0)$, being $(\rho,\phi)$ the polar coordinates in the Fourier domain $(u,v)$ and $B(\phi)$ an arbitrary azimuthal modulation. Therefore, the existence domain of any ND beam can also be extended by applying an appropriate quadratic phase factor to its complex amplitude.

Now we investigate the effects of transgressing the restriction $A\ll 1$. As first task, we compute the Fourier spectrum amplitudes of the PMGB beam defined in Eq. (13), by assuming $A=1, 2, 3$ and $4$. The computed Fourier spectrum amplitudes $|G(u,v)|$, displayed in Fig. 7, show evident shape deviations respect to the field amplitude $|g_0(x,y)|$ [in Fig. 4 (a)]. These deviations are due to the fact that the required restriction $A\ll 1$ is no longer fulfilled for the adopted $A$ values. In Fig. 8 we depict the normalized transverse amplitudes of the propagated PMBG beams for $A=3$ and $A=4$, at the propagation distance $z_0=6z_C$. As expected, the propagated PMBG beams tend to show the shapes of the Fourier spectrum amplitudes $|G(u,v)|$, respectively shown in Figs. 7 (c) and 7 (d).

The comparison of field amplitudes, made until now, is appropriate to enhance errors in small field levels. In the following computations we display the intensities of the fields, which enhance the similitude of low field values. In Fig. 9 (a) we display the intensity of the field $g_0(x,y)$ under discussion [which corresponds to the BG beam amplitude in Fig. 4 (a)]. On the other hand, in Figs. 9(b)-9(d) we show the intensities of the Fourier spectra of the PMBG beams, obtained with the parameters $A=1$, 2, and 3, respectively. The Fourier spectra intensities ($|G|^2$) in Figs. 9 provide (using the appropriate scaling) the intensity profiles of the far field propagated PMBG beams, for the different $A$ values. The far field intensities of the PMBG beams for $A=1$ and 2 [Fig. 9(b) and 9 (c)] present the features of the unmodified BG beam intensity, with high fidelity. The far field PMBG beam intensity for $A=3$ show clear errors beyond the first local minimum.

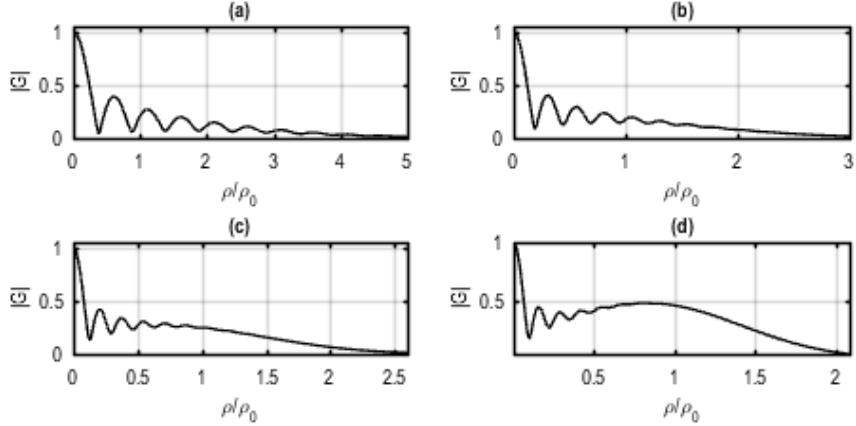

Fig. 7. Transverse amplitude of the Fourier transform of the PMBG field [Eq. (13)] obtained with $A=1$, 2, 3, and 4.

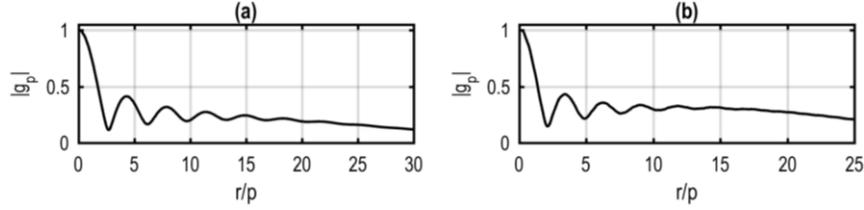

Fig. 8. Normalized transverse amplitudes of the propagated PMBG beams for $A=3$ and 4, at $z_0=6z_C$.

Even when the condition in Eq. (4) has been transgressed by assuming relatively high values of $A$, the transverse scale of the Fourier spectra in Fig. 9 is still proportional to $A^{-1}$ (or $R^{-1}$) in agreement with Eq. (5). It can also be verified that the high quality far field shape preservation, in the cases $A=1$ and $A=2$, also occurs for the corresponding near field propagated PMBG beams.

To evaluate the shape deviation between two intensity profiles $I_0(r)$ and $I(r)$ we employ the formula

$$E = \frac{\int_0^{r_p} r[I_0(r) - \alpha I(r)]^2 dr}{\int_0^{r_p} r[I_0(r)]^2 dr}. \tag{14}$$

We assume that the intensities $I_0(r)$ and $I(r)$ are defined in a circular domain of radius $r_p$ and that $\alpha$ is a constant that allows the best fitting between $I_0(r)$ and $I(r)$. Thus, this constant is obtained from the local minimum condition $\delta E/\delta\alpha=0$. Due to the factor $r$ in the integrands of Eq. (14), the integrals correspond to the circular 2-D domain. To measure the shape deviation of the propagated BG beam $g_{0p}(x,y)$, $I_0(r)$ and $I(r)$ are assumed as the transverse intensities of $g_0(x,y)$ and $g_{0p}(x,y)$ respectively. On the other hand, to measure the shape preservation of the propagated PMBG beam $g_p(x,y)$, $I(r)$ corresponds to the transverse intensity of $g_p(Mx,My)$.

The factor $M$ in this relation provides a compression in the field $g_p$ that reverts its scaling attained during propagation. This procedure allows a fair comparison of the PMBG beam intensity shapes at different propagation planes.

The computed intensity profile deviation $E$, versus the propagation distance $z_0$, for the propagated BG beam, is depicted in Fig. 10 (a) (black line). On the other hand, the deviation $E$, for different propagated PMBG beams (which are the phase modulated ones), is shown in the colored lines of Fig. 10 (a). In the latter case we have considered three values for the parameter $A$ (1/4, 1 and 2). In this figure, the shape error for the unmodified beam tends quickly to the maximum value (1) when the propagation distance tends to 1.5 $z_C$.

On the other hand, the shape errors for the modified beams are below 0.02 (2%) for the different $A$ values. Alternate error plots, which allow the display of the full error ranges in all the cases (in the employed propagation domain), are shown in Fig. 10 (b). In this case we displayed the natural logarithm of $E$.

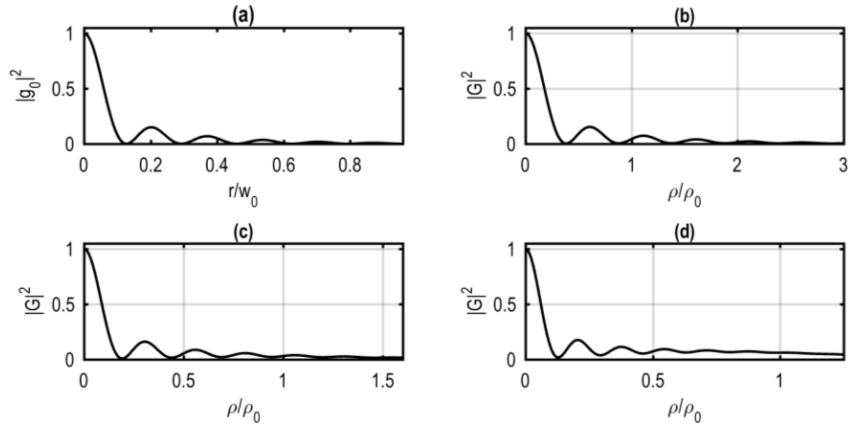

Fig. 9. Transverse intensities for (a) the BG beam and for the Fourier spectra of the PMBG beams, obtained with $A$ equal to (b) 1, (c) 2, and (d) 3.

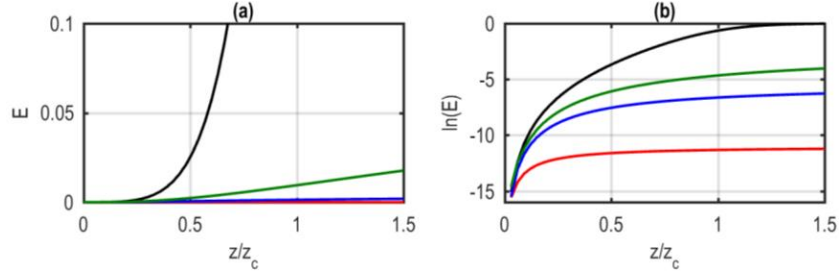

Fig. 10. Intensity deviations (a) $E$ and (b) $ln(E)$ for the propagated BG beam (black) and for the propagated PMBG beams with parameter $A$ equal to 1/4 (red), 1 (blue), and 2 (green).

## 4. Final remarks

We have proved that any optical field, whose Fourier spectrum is contained in a circle of finite radius ($\rho_C$), can be transformed into a beam with invariant shape during propagation, at the expense of a scaling in its transverse extension. The transformation consists in modulating the field by a quadratic phase factor, with a curvature radius R that fulfills the restriction $\lambda R \rho_C^2 \ll 1$. As a particular result, we proved that the Fourier transform of the phase modulated field inherits the mathematical form of this beam, except that the quadratic phase in the Fourier spectrum is conjugated. A straightforward method to modulate the original field $g_0(x,y)$ with a quadratic phase consists in transmitting it through a divergent lens.

The analyzed propagation invariance is an approximation whose errors decrease by reducing the curvature radius in the employed quadratic phase modulation. Specifically, the restriction $\lambda R \rho_C^2 \ll 1$ ensures almost negligible errors in the analyzed field invariance.

However, the results are not affected too much if the restriction $\lambda R \rho_C^2 = A \ll 1$ is moderately transgressed. We illustrated this result in Fig. 9, considering PMBG beams obtained with $A=1$, 2, and 3. The advantage of adopting relatively large $A$ values is that the quadratic phase modulation, that modifies the BG beam, presents relatively large curvature radii. As a consequence, the re-scaling in the transverse field extension, during propagation, reduces its growth speed.

A plausible conclusion is that the existence of a beam in an unbounded propagation range is only possible if one accepts a z-dependent re-scaling of its transverse extension. However, it is pending to develop a general proof of this assertion.

**References**

1. J. Durnin, J. J. Miceli, Jr., and J. H. Eberly, "Diffraction-free beams", Phys. Rev. Lett. **58**, 1499–1501 (1987).
2. McGloin, D. and Dholakia, K., "Bessel beams: diffraction in a new light," Contemporary Physics **46**, 15-28 (2005).
3. F. O. Fahrbach, P. Simon, and A. Rohrbach, "Microscopy with self-reconstructing beams," Nature Photonics **4**, 780-785 (2010).
4. F. Gori and G. Guattary, "Bessel-Gauss beams," Optics Commun. **64**, 491-495 (1987).
5. J. C. Gutiérrez-Vega, M. D. Iturbe-Castillo, and S. Chávez-Cerda, "Alternative formulation for invariant optical fields: Mathieu beams," Opt. Lett. **25**, 1493-1495 (2000).
6. Miguel A. Bandres, Julio C. Gutiérrez-Vega, and Sabino Chávez-Cerda, "Parabolic nondiffracting optical wave fields," Opt. Lett. **29**, 44-46 (2004).
7. S. N. Khonina, V. V. Kotlyar, V. A. Soifer, J. Lautanen, M. Honkanen, and J. Turunen, "Generating a couple of rotating nondiffracting beams using a binary-phase DOE," Optik-International Journal for Light and Electron Optics **110**, 137-144 (1999).
8. G. A. Siviloglou and D. N. Christodoulides. "Accelerating finite energy Airy beams," Optics Letters **32**, 979-981 (2007).
9. M.A. Bandres and J.C. Gutiérrez-Vega, "Airy-Gauss beams and their transformation by paraxial optical systems," Optics Express **15**, 16719-16728 (2007).
10. S.N. Khonina, "Specular and vortical Airy beams," Opt. Commun. **284**, 4263-4271 (2011).
11. A. E. Siegman, Lasers (Universe Science Books, 1986) pp. 642-652.
12. Z. Jaroszewicz and J. Morales, "Lens axicons: systems composed of a diverging aberrated lens and a perfect converging lens," J. Opt. Soc. Am. A **15**, 2383-2390 (1998).
13. T. Aruga, S. W. Li, S. Yoshikado, M. Takabe, and R. Li, "Nondiffracting narrow light beam with small atmospheric turbulence-influenced propagation," Appl. Opt. **38**, 3152-3156 (1999).
14. V. Belyi, A. Forbes, N. Kazak, N. Khilo, and P. Ropot, "Bessel-like beams with z–dependent cone angles," Opt. Express **18**, 1966-1973 (2010) .
15. Y Ismail, N Khilo, V Belyi, and A Forbes, "Shape invariant higher-order Bessel-like beams carrying orbital angular momentum," J. Opt. **14**, 1-12 (2012).
16. J. W. Goodman, Introduction to Fourier Optics, second Ed, (McGraw-Hill Companies, 1988) pp. 73-75.
17. M. J. Caola, "Self-Fourier functions," J. Phys. A: Math. Gen. **24**, L1143-L1144 (1991).
18. A. W. Lohmann and D. Mendlovic, "Self-Fourier objects and other self-transform objects," J. Opt. Soc. Am. A **9**, 2009-2012 (1992).
19. S. G. Lipson, "Self-Fourier objects and other self-transform objects: comment," J. Opt. Soc. Am. A **10**, 2088-2089 (1993).